\documentstyle[sprocl_ins,epsf]{article}

\def\be{\begin{equation}}
\def\ee{\end{equation}}
\def\bea{\begin{eqnarray}}
\def\eea{\end{eqnarray}}
\def\b{\begin{eqnarray*}}
\def\e{\end{eqnarray*}}

\begin{document}
\title{Instantons and Monopoles for Nonperturbative QCD } 
\author{H.~Suganuma, F.~Araki, M.~Fukushima, H.~Ichie,Y.~Koma,
       \\ S.~Sasaki, A.~Tanaka, H.~Toki and S.~Umisedo}
\address{Research Center for Nuclear Physics (RCNP), Osaka University\\
Mihogaoka 10-1, Ibaraki, Osaka 567-0047, Japan}

\maketitle\abstract{
We study the nonperturbative QCD in terms of 
topological objects, {\it i.e.}, QCD-monopoles and instantons. 
In the 't~Hooft abelian gauge, 
QCD is reduced into an abelian gauge theory 
with QCD-monopoles, and the confinement mechanism 
can be understood with the dual superconductor picture. 
Regarding the flux-tube ring as the glueball, 
we study the glueball properties by combining 
the dual Ginzburg-Landau theory and the Nambu-Goto action. 
We find the strong correlations between instantons and 
monopole world-lines 
both in the continuum theory and in the lattice QCD.
The nonperturbative QCD vacuum is filled with 
instantons and monopoles, 
and dense instantons generate a macroscopic network of 
the monopole world-line, which would be responsible for the 
confinement force in the infrared region. 
}

\section{Introduction}
Quantum chromodynamics (QCD) is the fundamental theory 
of the strong interaction 
[\cite{greiner}-\cite{cheng}].
In spite of the simple form of the QCD lagrangian,
$$
%{\cal L}_{\rm QCD}=-{1 \over 2} {\rm tr} G_{\mu \nu }G^{\mu \nu }
%+\bar q (\gslash D-m_q) q, 
{\cal L}_{\rm QCD}=-{1 \over 2} {\rm tr} G_{\mu \nu }G^{\mu \nu }
+\bar q (\not{D}-m_q) q, 
$$
it miraculously provides quite various phenomena 
like color confinement, dynamical chiral-symmetry breaking, 
non-trivial topologies, quantum anomalies and so on (Fig.1).
It would be interesting to compare QCD with the history of 
the Universe, because a quite simple `Big Bang' 
also created various things including galaxies, stars, lives and 
thinking reeds. 
Therefore, QCD can be regarded as an interesting 
miniature of the history of the Universe. 
This is the most attractive point of the QCD physics.

Since it is quite difficult to understand the various QCD phenomena and 
their underlying mechanism at the same time, many methods 
and models have been proposed to understand each phenomenon (Table 1).
We show in Fig.2 a brief sketch on the history of QCD and 
typical QCD effective models [\cite{greiner}]. 
In '80s, chiral symmetry breaking was the central issue. 
The chiral bag model, the NJL model and the $\sigma$ model were 
formulated with referring chiral symmetry.
In '90s, on the other hand, the confinement physics is providing 
an important current of the hadron physics, since 
recent lattice QCD studies 
[\cite{hioki}] 
shed a light on the confinement mechanism. 
The key word for the understanding of confinement 
is the ``duality'', which is recently paid attention by 
many theoretical particle physicists after Seiberg-Witten's discovery 
on the essential role of monopole condensation for the confinement 
in a supersymmetric version of QCD 
[\cite{seiberg}]. 
The origin of color confinement can be recognized 
as the dual Higgs mechanism by monopole condensation, 
and the nonperturbative QCD vacuum is regarded as 
the dual superconductor 
[\cite{giacomo}]. 
The dual Ginzburg-Landau (DGL) theory 
[\cite{suzuki}-\cite{umisedo}] 
was formulated from QCD with this picture, and provides a useful 
framework for the study of the nonperturbative QCD.

\section{Color Confinement and Dual Higgs Mechanism}
We briefly show the modern current of the confinement physics. 
The QCD vacuum can be regarded as the dual version of the 
superconductor, which was firstly pointed out by Nambu, 't~Hooft 
and Mandelstam 
[\cite{nambu}].
Here, the ``dual version'' means the interchange between 
the electric and magnetic sectors.
With referring Table 2, we compare the ordinary 
electromagnetic system, the superconductor and 
the nonperturbative QCD vacuum regarded as the dual superconductor. 

In the ordinary electromagnetism in the Coulomb phase, 
both electric flux and magnetic flux 
are conserved, respectively. The electric-flux conservation 
is guarantied by the ordinary gauge symmetry. 
On the other hand, the magnetic-flux conservation 
is originated from the dual gauge symmetry [\cite{suganumaA}-\cite{suganumaE}], 
which is the generalized version of the Bianchi identity. 
As for the inter-charge potential in the Coulomb phase, 
both electric and magnetic potentials are Coulomb-type. 

The superconductor in the Higgs phase is characterized by 
electric-charge condensation, which leads to the Higgs mechanism or 
spontaneous breaking of the ordinary gauge symmetry, 
and therefore the electric flux is no more conserved. 
In such a system obeying the London equation, 
the electric inter-charge potential becomes short-range Yukawa-type 
similarly in the electro-weak unified theory. 
On the other hand, the dual gauge symmetry is not broken, so that the 
magnetic flux is conserved, but is squeezed like a one-dimensional 
flux tube due to the Meissner effect.
As the result, the magnetic inter-charge potential becomes 
linearly rising like a condenser.

The nonperturbative QCD vacuum regarded as the dual Higgs phase 
is characterized by color-magnetic monopole condensation, 
which leads to the spontaneous breaking of the dual gauge symmetry. 
Therefore, color-magnetic flux is not conserved, 
and the magnetic inter-change potential becomes short-range Yukawa-type.
Note that the ordinary gauge symmetry is not broken by such monopole 
condensation. Therefore, color-electric flux is conserved, 
but is squeezed like a one-dimensional flux-tube or a string 
as a result of the dual Meissner effect.
Thus, the hadron flux-tube is formed in the monopole-condensed 
QCD vacuum, and the electric inter-charge potential 
becomes linearly rising, which confines the color-electric charges 
[\cite{suganumaA}-\cite{umisedo}]. 

As a remarkable fact in the duality physics, 
these are two ``see-saw relations'' [\cite{suganumaD}] 
between the electric and magnetic sectors. 
%\nextline
\\
(1) There appears the Dirac condition $eg=4\pi$ [\cite{suganumaA}] in QCD. 
Here, unit electric charge $e$ is the gauge coupling constant, 
and unit magnetic charge $g$ is the dual gauge coupling constant. 
Therefore, a strong-coupling system in one sector corresponds to 
a weak-coupling system in the other sector.
\\
(2) The long-range confinement system in one sector 
corresponds to a short-range (Yukawa-type) interaction system in the 
other sector.

Let us consider usefulness of the latter ``see-saw relation''.
One faces highly non-local properties among the color-electric 
charges in the QCD vacuum because of the long-range linear 
confinement potential.
Then, the direct formulation among the electric-charged variables 
would be difficult due to the non-locality, 
which seems to be a destiny in the long-distance confinement physics. 
However, one finds a short-range Yukawa potential 
in the magnetic sector, and therefore the electric-confinement 
system can be approximated by a local formulation 
among magnetic-charged variables. 
Thus, the confinement system, which seems highly non-local,  
can be described by a short-range interaction theory using 
the dual variables. This is the most attractive point 
in the dual Higgs theory.

Color-magnetic monopole condensation is necessary for 
color confinement in the dual Higgs theory. 
As for the appearance of color-magnetic monopoles in QCD, 't~Hooft 
proposed an interesting idea of the abelian gauge fixing
[\cite{thooft},\cite{ezawa}], which is defined by the diagonalization 
of a gauge-dependent variable. 
In this gauge, QCD is reduced into an abelian gauge theory with 
the color-magnetic monopole [\cite{suganumaA},\cite{thooft}], 
which will be called as QCD-monopoles hereafter. 
Similar to the 't~Hooft-Polyakov monopole 
[\cite{cheng}] in the Grand Unified theory 
(GUT), the QCD-monopole appears from a hedgehog configuration 
corresponding to the non-trivial homotopy group 
$\pi_2\{ {\rm SU}(N_c)/{\rm U}(1)^{N_c-1} \}=Z_\infty ^{N_c-1}$ 
on the nonabelian manifold. 

Many recent studies based on the lattice QCD 
in the 't~Hooft abelian gauge 
show QCD-monopole condensation in the confinement phase, 
and strongly support abelian dominance and monopole dominance 
for the nonperturbative QCD (NP-QCD), e.g., 
linear confinement potential, dynamical chiral-symmetry breaking 
(D$\chi $SB) and instantons 
[\cite{hioki},\cite{suganumaD},\cite{suganumaE},\cite{miyamura}-\cite{suganumaH}]. 
Here, abelian dominance means that QCD phenomena is described only 
by abelian variables in the abelian gauge. 
Monopole dominance is more strict, and means that 
the essence of NP-QCD is described only by the singular 
monopole part of abelian variables 
[\cite{hioki},\cite{suganumaD},\cite{suganumaE},\cite{miyamura}-\cite{suganumaH}].

Figure 3 is the schematic explanation on abelian dominance 
and monopole dominance observed in the lattice QCD. 

\noindent
(a) Without gauge fixing, it is difficult to extract 
relevant degrees of freedom in NP-QCD. 

\noindent
(b) In the 't~Hooft abelian gauge, 
QCD is reduced into an abelian gauge theory 
including the electric current $j_\mu $ and the magnetic current $k_\mu $.
Particularly in the maximally abelian (MA) gauge 
[\cite{hioki},\cite{miyamura}-\cite{suganumaH}], 
only U(1) gauge degrees of freedom including monopole is relevant for 
NP-QCD, while off-diagonal components do not contribute to NP-QCD: 
abelian dominance.
Here, the very long monopole world-line appears 
in the confinement phase. 
We show the lattice data of the monopole world-line in Fig.4. 

\noindent
(c) The U(1)-variable can be decomposed into the regular 
photon part and the singular monopole part 
[\cite{miyamura}-\cite{suganumaH},\cite{dgt}], which corresponds to 
the separation of $j_\mu$ and $k_\mu$. 
The monopole part leads to NP-QCD 
(confinement, D$\chi $SB, instanton): monopole dominance.
On the other hand, the photon part is almost trivial similar to 
the QED system.

\noindent
Thus, monopoles in the MA gauge can be 
regarded as the relevant collective mode for NP-QCD, 
and therefore the NP-QCD vacuum can be identified as 
the dual-superconductor in a realistic sense.

\section{Glueballs in the Dual Ginzburg-Landau Theory}

In the dual Higgs theory, the monopole appears as 
a scalar glueball with $J^{PC}=0^{++}$ like the Higgs particle 
in the standard model [\cite{cheng}]. 
Here, the glueball is an elementary excitation without valence 
quarks in the NP-QCD vacuum, and is considered as the 
{\it flux-tube ring} in the DGL theory, 
while ordinary hadrons are described as 
open flux-tubes with the valence quarks at the ends. 
In this chapter, we study the glueball properties by analyzing 
the flux-tube ring solution in the DGL theory.

First, we calculate the energy $E_{\rm cl}(R)$ of the 
classical ring solution with the radius $R$ in the DGL theory, 
and estimate the effective string tension $\sigma _{\rm eff}(R)$ satisfying 
$E_{\rm cl}(R)=2\pi R \sigma _{\rm eff}(R)$. 
Although the flux-tube ring seems to shrink at the classical level, 
this ring solution is stabilized against such a collapse 
by the quantum effect.

Second, regarding the flux-tube ring as a 
{\it relativistic closed string} 
with the effective string tension $\sigma _{\rm eff}(R)$, 
we introduce the quantum effect to the ring solution using the 
{\it Nambu-Goto (NG) action} in the string theory 
>From the Hamiltonian of the NG action, we find 
the energy of the closed string as 
$$
E(R,P_R)=\sqrt{P_R^2+ \{ 2\pi R \sigma_{\rm eff}(R) \}^2},
$$ 
where $P_R$ is the canonical conjugate momentum of the coordinate $R$.

Using the uncertainty relation $P_R \cdot  R \geq 1$ as the 
quantum effect, 
the energy and radius of the ring solution are estimated as 
$M \simeq$ 1.2 GeV and $R \simeq$ 0.22 fm, respectively. 
Although the angular momentum projection 
related to the ring motion is needed 
to extract the physical glueball state with definite quantum numbers, 
the DGL theory can provide a useful method for the study 
of the glueball.

\section{Correlation between Instanton and QCD-monopole}
In the Euclidean metric, 
the ``instanton'' appears as the {\it classical solution of QCD} 
with the {\it nontrivial topology} corresponding 
to $\pi_{3}\{ {\rm SU}(N_c) \}$ =$Z_\infty $, 
and is responsible for the U$_{\rm A}$(1) anomaly or 
the large mass generation of $\eta $' 
[\cite{diakonov},\cite{shuryak}]. 

Recent lattice studies [\cite{hioki},\cite{miyamura}-\cite{suganumaH}] 
indicate {\it abelian dominance} for nonperturbative quantities 
in the maximally abelian (MA) gauge.
{\it In the abelian-dominant system, 
the instanton seems to lose the topological basis 
for its existence, and therefore it seems unable to 
survive in the abelian manifold. 
However, even in the MA gauge, nonabelian components remain 
relatively large around the topological defect, {\it i.e.} monopoles, 
and therefore instantons are expected to survive only around the 
monopole world-lines in the abelian-dominant system} 
[\cite{suganumaC},\cite{suganumaD},\cite{suganumaE},\cite{suganumaF}-\cite{suganumaH}].

We have pointed out such a close relation between instantons 
and monopoles, and have demonstrated it in the continuum QCD with 
$N_c$=2 using the MA gauge 
[\cite{fukushima}] or 
the Polyakov-like gauge, where $A_4(x)$ is diagonalized 
[\cite{suganumaC},\cite{suganumaD},\cite{suganumaE},\cite{suganumaF}-\cite{suganumaH}] 

We summarize our previous analytical works as follows. 
%\item{\rm 
(1) 
Each instanton accompanies a monopole current.
In other words, instantons only live along the monopole world-line. 
%\item{\rm 
(2) 
The monopole world-line is unstable against a small fluctuation 
of the location or the size of instantons,  although it is relatively 
stable inside the instanton profile. 
%\item{\rm 
(3) 
In the multi-instanton system, monopole world-lines 
become highly complicated, 
and there appears a huge monopole clustering, which 
covers over the whole system, for the dense instanton system. 
%\item{\rm 
(4) 
At a high temperature, the monopole world-lines are drastically 
changed, and become simple lines along 
the temporal direction.

We study also the correlation between instantons and monopoles 
in the MA gauge using the {\it Monte Carlo simulation} in the 
{\it SU(2) lattice gauge theory} [\cite{suganumaD},\cite{suganumaE},
\cite{suganumaF}-\cite{suganumaH}].
The SU(2) link variable can be separated into the singular 
monopole part and the regular photon part as shown in Fig.3. 
Using the cooling method, we measure the topological quantities 
($Q$ and $I_Q$) in the monopole and photon sectors as well as in the 
ordinary SU(2) sector. 
Here, $I_Q \equiv \int d^4x |{\rm tr}(G_{\mu\nu} \tilde G_{\mu\nu})|$ 
corresponds to the total number $N_{\rm tot}$ of instantons and 
anti-instantons.

On the $16^4$ lattice with $\beta=2.4$, we find that instantons exist only in 
the monopole part in the MA gauge, which means monopole dominance 
for the topological charge [\cite{miyamura}-\cite{suganumaH}].
Hence, monopole dominance is also expected for the U$_{\rm A}$(1) 
anomaly and the large $\eta'$ mass.

We study the finite-temperature system using the 
$16^3 \times 4$ lattice with various $\beta$ around 
$\beta_c \simeq 2.3$ [\cite{suganumaD},\cite{suganumaH}]. 
We show in Fig.5 the correlation between $I_Q({\rm SU(2)})$ 
and $I_Q({\rm Ds})$, which are measured 
in the SU(2) and monopole sectors, respectively, after 50 cooling sweeps.
{\it The monopole part holds the dominant topological charge 
in the full SU(2) gauge configuration.} 
On the other hand, $I_Q(Ph)$, measured in the photon part, 
vanishes quickly by several cooling sweeps. 
Thus, {\it monopole dominance for instantons is observed in the 
confinement phase even at finite temperatures} [\cite{suganumaD},\cite{suganumaH}].
Near the critical temperature $\beta_c \simeq 2.3$, a large 
reduction of $I_Q$ is observed. In the deconfinement phase, 
$I_Q$ vanishes quickly by several cooling sweeps, 
which means the absence of the instanton, in the SU(2) and 
monopole sectors as well as in the photon sector [\cite{suganumaD},\cite{suganumaH}].
Therefore, the gauge configuration becomes similar to the photon part 
in the deconfinement phase.

Finally, we study the correlation between the instanton number and 
the monopole loop length. Our analytical studies suggest the appearance 
of the highly complicated monopole world-line 
in the multi-instanton system 
[\cite{suganumaD},\cite{suganumaH},\cite{fukushima}]. 
We conjecture that the existence of instantons promotes monopole 
condensation, which is characterized by a long complicated 
monopole world-line covering over ${\bf R}^4$ 
observed in the lattice QCD [\cite{hioki},\cite{suganumaD}]. 
Using the lattice QCD, we measure the total monopole-loop length $L$ 
and the integral of the absolute value of the topological density 
$I_Q$, which corresponds to the total number $N_{\rm tot}$ of instantons and anti-instantons.
We plot in Fig.6 the correlation between $I_Q$ and $L$ in the MA gauge 
after 3 cooling sweeps on the $16^3 \times 4 $ lattice with various $\beta$.
A {\it linear correlation} is clearly found between $I_Q$ and $L$, 
although naive dimension counting suggests 
$I_Q^{1/4} = {\rm const} \cdot L^{1/3}$. 

{\it In terms of the microscopic correlation, 
each instanton accompanies a small monopole loop nearby,} 
whose length would be proportional to the instanton size 
[\cite{suganumaH},\cite{fukushima}-\cite{markum}].
Since instantons are dense in the confinement phase, 
monopole loops around them overlap each other, 
and there appears {\it a macroscopic network of the monopole world-line,} 
which bonds neighboring instantons [\cite{fukushima}].
In fact, {\it the NP-QCD vacuum is filled with 
two sort of topological objects, instantons and monopoles, 
and dense instantons generate the macroscopic 
monopole clustering, which would be responsible for the 
confinement force in the infrared region.} 

%\refout
\section{References*}

%\endpage
							
Fig.1 : A brief sketch of the QCD physics.

Fig.2 : A brief history QCD and QCD effective models.

Fig.3 : Abelian dominance and monopole dominance: 
\\ %\nextline
(a) Without gauge fixing, it is difficult to extract 
relevant degrees of freedom in NP-QCD. 
\\ %\nextline
(b) In the MA gauge, 
QCD is reduced into an abelian gauge theory 
including the electric current $j_\mu $ and the magnetic current $k_\mu $.
Only U(1) gauge degrees of freedom including monopole is relevant for 
NP-QCD, while off-diagonal components do not contribute to NP-QCD: 
abelian dominance.
Here, the very long monopole world-line appears 
in the confinement phase. 
\\ %\nextline 
(c) The U(1)-variable can be decomposed into the regular 
photon part and the singular monopole part, 
which corresponds to 
the separation of $j_\mu$ and $k_\mu$. 
The monopole part leads to NP-QCD 
(confinement, D$\chi $SB, instanton): monopole dominance.
On the other hand, the photon part is almost trivial similar to 
the QED system.

Fig.4: The monopole world-lines projected on 
the 3-dimensional space 
in the confinement phase at $\beta $=2.2 
in the SU(2) lattice QCD with $16^3 \times 4$.

Fig.5: The correlation between $I_Q$(SU(2)) and $I_Q$(DS) 
in the MA gauge after 50 cooling sweeps on the $16^3\times 4$ lattice; 
$\circ$ denotes in the confinement phase ($\beta $=2.2), 
$\times$ at the critical temperature ($\beta $=2.3) and 
$\triangle$ in the deconfinement phase ($\beta $=2.35). 
The monopole part holds instantons 
in the full SU(2) gauge configuration in the confinement phase, 
while there is no instanton in the SU(2) and monopole sectors 
in the deconfinement phase similarly in the photon sector.

Fig.6: The SU(2) lattice QCD data for the correlation between the total 
monopole-loop length $L$ 
and $I_Q$ (the total number of instantons and anti-instantons) 
in the MA gauge.
We plot the data at 3 cooling sweep on the $16^3 \times 4$ lattice 
with various $\beta$. 

Table 1 : Features of the lattice QCD and the QCD effective models

Table 2 : The comparison among the ordinary electromagnetic system, the 
superconductor and the nonperturbative QCD vacuum regarded as the 
QCD-monopole condensed system.

\end{document}